\documentclass[noend]{llncs}
\usepackage{amsmath}  
\usepackage{semantic}
\usepackage{amssymb}
\usepackage{xspace}
\usepackage{colonequals}
\usepackage[linesnumbered]{algorithm2e}
\usepackage{multicol}
\usepackage{subfig}
\usepackage{graphicx}

\SetNlSty{tiny}{}{\hspace{1.5em}}

\newcommand{\eg}{e.g.\xspace}
\newcommand{\ie}{i.e.\xspace}

\newcommand{\shexone}{ShEx~1.0\xspace}
\newcommand{\shextwo}{ShEx~2.0\xspace}

\newcommand{\nodes}{\mathrm{\mathbf{N}odes}}
\newcommand{\Blank}{\ensuremath{\mathsf{Blank}\xspace}}
\newcommand{\graph}{\mathrm{\mathbf{G}}}

\newcommand{\IRI}{\ensuremath{\mathsf{IRI}\xspace}}
\newcommand{\Lit}{\ensuremath{\mathsf{Lit}\xspace}}
\newcommand{\blank}{\ensuremath{\mathsf{\_b}\xspace}}

\newcommand{\sch}{\ensuremath{\mathbf{S}}}
\newcommand{\SLabels}{\ensuremath{\mathbf{L}\xspace}}
\newcommand{\ddef}{\ensuremath{\mathbf{def}\xspace}}

\newcommand{\ShExpr}{\ensuremath{\mathsf{ShExpr}\xspace}}
\newcommand{\ValueDescr}{\ensuremath{\mathsf{ValueDescr}\xspace}}
 
\newcommand{\NeighbDescr}{\ensuremath{\mathsf{NeigDescr}\xspace}}
\newcommand{\TExpr}{\ensuremath{\mathsf{TExpr}\xspace}}

\newcommand{\ShapeAnd}{\ensuremath{\mathsf{ShapeAnd}\xspace}}
\newcommand{\ShapeOr}{\ensuremath{\mathsf{ShapeOr}\xspace}}
\newcommand{\ShapeNot}{\ensuremath{\mathsf{ShapeNot}\xspace}}

\newcommand{\AND}{\ensuremath{\mathbin{\mathsf{AND}}}}
\newcommand{\OR}{\ensuremath{\mathbin{\mathsf{OR}}}}
\newcommand{\NOT}{\ensuremath{\mathop{\mathsf{NOT}}}}

\newcommand{\TriplePattern}{\ensuremath{\mathsf{TriplePattern}}\xspace}
\newcommand{\SomeOfExpr}{\ensuremath{\mathsf{SomeOfExpr}}\xspace}
\newcommand{\EachOfExpr}{\ensuremath{\mathsf{EachOfExpr}}\xspace} 
\newcommand{\RepetExpr}{\ensuremath{\mathsf{RepetExpr}}\xspace}
\newcommand{\EMPTY}{\ensuremath{\mathsf{EMPTY}\xspace}}

\newcommand{\PropSet}{\ensuremath{\mathsf{PropSet}\xspace}}
\newcommand{\ShapeRef}{\ensuremath{\mathsf{ShapeRef}\xspace}}

\newcommand{\typing}{\mathit{typing}}
\newcommand{\sevaluedescr}{se-value-descr\xspace}
\newcommand{\seneighbdescr}{se-neig-descr\xspace}
\newcommand{\seshapeand}{se-shape-and\xspace}
\newcommand{\seshapeor}{se-shape-or\xspace}
\newcommand{\seshapenot}{se-shape-not\xspace}
\newcommand{\tetriplepattern}{te-tpattern\xspace}
\newcommand{\tesomeof}{te-some-of\xspace}
\newcommand{\teeachof}{te-each-of\xspace}
\newcommand{\terepet}{te-repet\xspace}
\newcommand{\teempty}{te-empty\xspace}

\newcommand{\sat}{\vdash}
\newcommand{\match}{\vDash}

\newcommand{\neigh}{\ensuremath{\mathit{neigh}\xspace}}

\newcommand{\deppos}{\mathit{dep}^{+}}
\newcommand{\depneg}{\mathit{dep}^{-}}
\newcommand{\deps}{\mathit{dep}}

\newcommand{\strat}{\mathop{\mathit{strat}}}

\newcommand{\Typing}{\mathop{\mathit{Typing}}}

\newcommand{\true}{\mathit{true}}
\newcommand{\false}{\mathit{false}}

\newcommand{\recalgo}{\ensuremath{\mathit{prove}}}
\newcommand{\refinealgo}{\ensuremath{\mathit{refine}}}

\newcommand{\Dep}{\ensuremath{\mathit{Dep}}}

\newcommand{\Hyp}{\ensuremath{\mathit{Hyp}}}

\newcommand{\push}{\mathop{\mathit{push}}}
\newcommand{\pop}{\mathop{\mathit{pop}}}

\newenvironment{code}{\sf}{}
\newcommand{\codeil}[1]{{\textsf{#1}}}

\newcommand{\MQ}{\textsc{shapes\_sem}}

\newcommand{\ex}[1]{{\sf \textless{}#1\textgreater}}

\newcommand{\shacl}{\textsc{shacl}\xspace}
\newcommand{\sparql}{\textsc{sparql}\xspace}
\newcommand{\rdfschema}{RDF Schema\xspace}
\newcommand{\xmlschema}{XML Schema\xspace}
\newcommand{\rdf}{\textsc{rdf}\xspace}
\newcommand{\owl}{\textsc{owl}\xspace}
\newcommand{\foaf}{\textsc{foaf}\xspace}
\newcommand{\xml}{\textsc{xml}\xspace}
\newcommand{\iri}{\textsc{iri}\xspace}

\begin{document}

\title{Semantics and Validation of Shapes Schemas for RDF}

\author{Iovka Boneva\inst{1} \and Jose E. Labra Gayo\inst{2} \and Eric G. Prud'hommeaux\inst{3}}
\institute{
Univ. Lille - CRIStAL - F-59000 Lille, France.
\email{iovka.boneva@univ-lille.fr}
\and
University of Oviedo, Spain.
\email{labra@uniovi.es}
\and
W3C and Stata Center, MIT.
\email{eric@w3.org}
}
\maketitle

%%% Local Variables: 
%%% mode: latex
%%% TeX-master: "paper" 
%%% End: 

\begin{abstract}
We present a formal semantics and proof of soundness for shapes schemas, an expressive schema language for RDF graphs that is the foundation of Shape Expressions Language 2.0.
It can be used to describe the vocabulary and the structure of an RDF graph, and to constrain the admissible properties and values for nodes in that graph.
The language defines a typing mechanism called shapes against which nodes of the graph can be checked.
It includes an algebraic grouping operator, a choice operator and cardinality constraints for the number of allowed occurrences of a property.
Shapes can be combined using Boolean operators, and can use possibly recursive references to other shapes.

We describe the syntax of the language and define its semantics.
The semantics is proven to be well-defined for schemas that satisfy a reasonable syntactic restriction, namely stratified use of negation and recursion.
We present two algorithms for the validation of an RDF graph against a shapes schema.
The first algorithm is a direct implementation of the semantics, whereas the second is a non-trivial improvement.
We also briefly give implementation guidelines.
\end{abstract}

%%% Local Variables: 
%%% mode: latex
%%% TeX-master: "paper"
%%% End: 

\section{Introduction}
\label{sec:intro}
\rdf's distributed graph model encouraged adoption for publication and manipulation of \eg{} social and biological data.
Coding errors in data stores like DBpedia have largely been handled in a piecemeal fashion with no formal mechanism for detecting or describing schema violations.
Extending uptake into environments like medicine, business and banking requires structural validation analogous to what is available in relational or \xml{} schemas.

While \owl ontologies can be used for limited structural validation, they are generally used for formal models of reusable classes and predicates describing objects in some domain.
Applications typically consume and produce graphs composed of precise compositions of such ontologies.
A company's human resources records may leverage terms from \textsc{foaf} and Dublin Core, but only certain terms, composed into specific structures, and subject to additional use-specific constraints.
We would no more want to impose the constraints of a single human resources application suite on \textsc{foaf} and Dublin Core than we would want to assert that such applications need to consume all ontologically valid permutations of \textsc{foaf} and Dublin Core entities.
Further, open-world constraints on \owl ontologies make it impossible to use conventional \owl tools to \eg{} detect missing properties.
Shape expression schemas (\shexone) \cite{semantics2014,icdt2015} were introduced as a high level language in which it is easy to mix terms from arbitrary ontologies.
They provide a schema language in which one can define structural constraints (arc labels, cardinalities, datatypes, etc.) and since version~2.0 (\shextwo)\footnote{Shape Expressions Language 2.0. \url{http://shex.io/shex-semantics/index.html}}, mix them using Boolean connectives (disjunction, conjunction and negation).

A schema language for any data format has several uses: 
communicating to humans and machines the form of input/output data; 
enabling machine-verification of data for production, publication, or consumption;
driving query and input interfaces;  static analysis of queries.
In this, ShEx provides a similar role as relational and \xml{} schemas.
A ShEx schema validates nodes in a graph against a schema construct called a \emph{shape}.
In \xml, validating an element against an XML Schema\footnote{W3C XML Schema. \url{http://www.w3.org/XML/Schema}} type or element or Relax NG\footnote{RELAX NG home page. \url{http://relaxng.org}} production recursively tests nested elements against constituent rules.
In ShEx, validating a node in a graph against a shape recursively tests the nodes which are the object of triples constrained in that shape.
An essential difference however is that unlike trees, graphs can have cycles and recursive definitions can yield infinite computation.
Moreover, \shextwo includes a negation operator, and it is well known that mixing recursion with negation can lead to incoherent semantics.

\paragraph{Contributions.}
In this paper we present \emph{shapes schemas}, a schema language that is the foundation of \shextwo (Sect.~\ref{sec:shapes}).
The precise relationship between shapes schemas and \shextwo is given at the end of Sect.~\ref{sec:shapes}.
We formally define the semantics of shapes schemas and show that it is sound for schemas that mix recursion and negation in a stratified manner (Sect.~\ref{sec:semantics}).
We then propose two algorithms for validating an RDF graph node against a shapes schema.
Both algorithms are shown to be correct w.r.t. the semantics (Sect.~\ref{sec:compute}).
We finally discuss future research directions and conclude (Sect.~\ref{sec:concl}).

\paragraph{Related Work.}
In \cite{icdt2015} we gave semantics for \shexone.
The latter does not use Boolean operators end because of negation, the extension to \shextwo (and thus to shapes schemas) is non trivial.

Closest to shapes schemas is the \shacl\footnote{Shapes Constraint Language (SHACL). \url{https://www.w3.org/TR/shacl/}} language both in terms of purpose and expressiveness.
\shacl also defines named constraints called shapes to be checked on \rdf graph nodes.
Unlike ShEx, \shacl is not completely independent from the \rdfschema vocabulary: {\sf rdfs:Class}es play a particular role there as a shape can be required to hold for all the nodes that are instances of some {\sf rdfs:Class}.
Therefore validation in \shacl requires partial \rdfschema entailment in order to discover all {\sf rdfs:Class}es of a node.
Regarding expressiveness, the main differences between \shacl and shapes schemas are that \shacl allows to define constraints based on property paths and for comparison of values; \shacl does not have the algebraic operators some-of and each-of and uses Boolean connectives for defining complex shapes; finally \shacl does not define the semantics of recursive shapes.

Ontology languages such as \owl, description logics or \rdfschema are not meant to define (complex) constraints on the data and we do not compare shapes schemas with them.
Proposals were made for using \owl with a closed world assumption in order to express integrity constraints \cite{Tiao10,Motik07}. They associate alternative semantics with the existing \owl syntax and can be misleading for users.

Some approaches use \sparql{} to express constraints on graphs (\textsc{spin}\footnote{SPIN - Modeling Vocabulary. \url{http://www.w3.org/Submission/spin-modeling/}}, RDFUnit \cite{kontokostasDatabugger}), or compile a domain specific language into \sparql{} queries~\cite{Fischer2015}.
\sparql{} allows to express complex constraints but does not support recursion.
While \sparql{} constraints can be validated by standard \sparql{} engines, they are har\-der to write and maintain compared to high-level schemas like ShEx and \shacl.

Description Set Profiles\footnote{Description Set Profiles: A constraint language for Dublin Core Application Profiles. \url{http://dublincore.org/documents/dc-dsp/}} is a constraint language that uses an \rdf{} vocabulary to define templates and constrain the value and cardinality of properties.
It does not have any equivalent of the each-of algebraic operator, and was not designed to be human-readable.

%%% Local Variables: 
%%% mode: latex
%%% TeX-master: "paper"
%%% End: 

\paragraph{Introductory Example.}
\label{sec:prelim}
Let {\sf is:} be a namespace prefix from some ontology, {\sf ex:} be the prefix used in the example schema and instance, and {\sf foaf:} and {\sf xsd:} be the standard \foaf and \textsc{xsd} prefixes, respectively.
The schema $\sch_0$ is as follows
{
  \vspace{-0.5cm}
  \small
  \begin{code}
    \begin{tabbing}
    \ex{ClientAndUser} \= $\to$ \= \hspace{7,2cm} \kill \\
    \ex{UserShape} \> $\to$  foaf:name @\ex{StringValue} ; foaf:mbox @\ex{IRIValue} [0;1]\\
    \ex{ProgShape} \> $\to$  ex:expertise @\ex{IRIValue} [0;*] ; ex:experience @\ex{ExpValueSet}\\
    \ex{ClientShape} \> $\to$ ex:clientNbr @\ex{IntValue} $\mid$ ex:clientAffil @\ex{AnythgShape}\\
    \ex{IssueShape} \> $\to$ is:reportedBy @\ex{ClientAndUser} ;\\
    \>\> is:reproducedBy @\ex{ProgShape} [1;5] ; \\
    \>\> is:relatedTo @\ex{IssueShape} [0;*] \\
    \ex{AnythgShape} \> $\to$ IRI @\ex{AnythgShape} [0;*]\\
    \ex{ClientAndUser} \> $\to$ $\mathit{Def}_{\text{Client}}$ ; IRI $\!-\!$ \{ex:clientNbr, ex:clientAffil\} @\ex{AnythgShape} [0;*]\\
\>\> AND $\mathit{Def}_{\text{User}}$ ; IRI $\!-\!$ \{foaf:name, foaf:mbox\} @\ex{AnythgShape} [0;*]\\
    \ex{StringValue} \> $\to$ xsd:string\\
    \ex{IRIValue} \> $\to$ IRI \\
    \ex{ExpValueSet} \> $\to$ \{ex:senior, ex:junior\}\\
    \ex{IntValue} \> $\to$ xsd:integer
  \end{tabbing}
  \end{code}
}
\noindent
where $\mathit{Def}_{\text{\sf Client}}$ is the definition of \ex{ClientShape}, and similarly for $\mathit{Def}_{\text{\sf User}}$, and $-$ in the definition of \ex{ClientAndUser} is the set difference operator.
The schema $\sch_0$ defines four shapes intended to describe users, programmers, clients and issues, respectively.
\ex{UserShape} requires that a node has one {\sf foaf:name} property with string value, and an optional {\sf foaf:mbox} that is an \iri.
The optional mailbox is specified by the cardinality constraint {\sf [0;1]}.
Other cardinality constraints used in $\sch_0$ are {\sf [0;*]} for zero or more, and {\sf [1;5]} for one up to five.
When no cardinality is given, the default is ``exactly one".
A \ex{ProgShape} node has zero or more {\sf ex:expertise} properties with values that are \iri{}s, and one {\sf ex:experience} property whose value is one among {\sf ex:senior} and {\sf ex:junior}.
A \ex{ClientShape} has either a {\sf ex:clientNbr} that is an integer, or a {\sf ex:clientAffil}(iation) with unconstrained value (\ie{} \ex{AnythgShape}), but not both.
Finally, an issue (\ex{IssueShape}) is reported by somebody who is client and user, is reproduced by one to five programmers, and can be related to zero or more issues.

The shapes in $\sch_0$ whose name contains {\sf Value} specify the set of allowed values for a node.
This can be the set of all values of some literal datatype (\eg{} string, integer), the set of all nodes of some kind (\eg{} \iri), or an explicitly given set (\eg{} \ex{ExpValueSet}).
\ex{AnythgShape} is satisfied by every node.
It states that the node can have zero or more outgoing triples whose predicates can be any \iri{}, and whose objects match \ex{AnythgShape}.
Finally, \ex{ClientAndUser} uses a conjunction to require that a node has both the client and the user properties.
Its definition is a bit technical.
The right hand side of the conjunction states that the node must have a {\sf foaf:name} and an optional {\sf foaf:mbox} ($\mathit{Def}_{\text{\sf User}}$).
Moreover (the ; operator), the node can have any number ({\sf [0;*]}) of properties that can be any \iri except for {\sf foaf:name} and {\sf foaf:mbox} and whose value is unconstrained.
The latter is necessary in order to allow the ``client'' properties required by the left hand side of the conjunction.

Graph $\graph_0$ here after is described by schema $\sch_0$.
Nodes \codeil{ex:issue1} and \codeil{ex:issue2} have shape \ex{IssueShape};
\codeil{ex:fatima} and \codeil{ex:emin} are \ex{ClientAndUser};
\codeil{ex:ren} and \codeil{ex:noa} have shape \ex{ProgShape}.

{
  \vspace{-0.3cm}
  \small
  \begin{code}
    \begin{multicols}{2}
    \begin {tabbing}
      \hspace{0.3cm} \= \hspace{4,1cm} \= \hspace{0.5cm} \= \hspace{0.2cm} \= \hspace{2cm} \= \kill
      ex:issue1 \\
      \> is:reportedBy ex:fatima ; \\
      \> is:reproducedBy ex:ren , ex:noa ; \\
      \> is:relatedTo ex:issue2 .\\
      ex:issue2 \\
      \> is:reportedBy ex:emin ; \\
      \> is:reproducedBy ex:ren ; \\
      \> is:relatedTo ex:issue1 .\\
      ex:fatima ex:clientNbr 1 ; \\
      \> foaf:name ``Fatima Smith''. \\
      ex:ren ex:expertise ex:semweb ; \\
      \> ex:experience ex:senior . \\
      ex:noa ex:experience ex:junior . \\
      ex:emin ex:clientAffil ``ABC''; \\
      \> foaf:name ``Emin V. Petrov'' ; \\
      \> foaf:mbox $<$mailto:evp@example.org$>$ .
    \end{tabbing}
  \end{multicols}
  \end{code}  
}

\paragraph{The RDF Graph Model.}
As usual, we assume three disjoint sets: $\IRI$ a set of \iri{}s, $\Lit$ a set of literals, and $\Blank$ a set of blank nodes.
An \emph{RDF graph} is a set of triples over $\IRI \cup \Blank \times \IRI \times \IRI \cup \Lit \cup \Blank$.
For a triple $(s, p, o)$ in some graph, $s$ is called its subject, $p$ is called its predicate, and $o$ is called its object.
We denote $\nodes(\graph)$ the set of \emph{nodes} of the graph $\graph$, that is, the elements that appear in a subject or object position in some triple of $\graph$.
The \emph{neighbourhood} of node $n$ in graph $\graph$ is the set of triples in $\graph$ that have $n$ as subject, and is denoted $\neigh_\graph(n)$ or simply $\neigh(n)$ when $\graph$ is clear from the context.
We use disjoint union on sets of triples, denoted $\uplus$: if $N, N_1, N_2$ are sets of triples, $N = N_1 \uplus N_2$ means that $N_1 \cup N_2 = N$ and $N_1 \cap N_2 = \emptyset$.

%%% Local Variables: 
%%% mode: latex
%%% TeX-master: "paper"
%%% End: 

\section{Shapes Schemas}
\label{sec:shapes}
A shapes schema $\sch$ defines a set of named shapes.
A shape is a description of the graph structure that can be visited starting from a particular node.
It can talk about the value of the node itself and about its neighbourhood.
Shapes can use (Boolean combinations of) other shapes and can be recursive.

Formally, a shapes schema $\sch$ is a pair $(\SLabels, \ddef)$, where $\SLabels$ is a set of \emph{shape labels} used as names of shapes and $\ddef$ is a function that with every shape label associates a shape expression.
In examples, we write $L \to S$ as short for $\ddef(L) = S$ (for a shape label $L$ and a shape expression $S$).

\paragraph{Shape Expressions.}
\label{sec:shape-expressions}
The grammar for shape expressions is given on Fig.~\ref{fig:grammar-shapes}.
A \emph{shape expression} (\ShExpr) is a Boolean combination of two atomic components: value description and neighbourhood description.
A neighbourhood description (\NeighbDescr) defines the expected neighbourhood of a node and is given by a triple expression (\TExpr, see below).
A value description (\ValueDescr) is a set that declares the admissible values for a node.
The set can contain \iri{}s, literals, and the special constant $\blank$ to indicate that the node can be a blank node.
\shextwo proposes concrete syntax for different kinds of value description sets (literal datatypes, regex patterns to be matched by \IRI{}s, intervals, etc.).
Here we focus on defining the semantics so the concrete syntax for such sets is irrelevant.
A \ValueDescr{} can be an arbitrary set with  the unique assumption that it has a finite representation for which membership can be effectively computed.
\begin{figure}[t]
\subfloat[Shape expressions.]{\label{fig:grammar-shapes}%
  \begin{minipage}[b]{0.45\linewidth}
    \begin{tabbing}
      \ValueDescr{} \= \hspace{0.5em} \=  \kill
      \ShExpr{} \> $\coloncolonequals$  \ValueDescr{} $\mid$ \NeighbDescr \\
      \>\> $\mid$ \ShapeAnd{} $\mid$ \ShapeOr{} \\
      \>\> $\mid$ \ShapeNot\\
      \ValueDescr{} \> $\coloncolonequals$ a subset of $\IRI \cup \Lit \cup \{\blank\}$ \\
      \NeighbDescr{} \> $\coloncolonequals$ \TExpr\\
      \ShapeAnd{} \> $\coloncolonequals$ \ShExpr{} {\sf `\AND'} \ShExpr \\
      \ShapeOr{} \> $\coloncolonequals$ \ShExpr{} {\sf `\OR'} \ShExpr \\
      \ShapeNot{} \> $\coloncolonequals$ {\sf `\NOT'} \ShExpr
    \end{tabbing}
  \end{minipage}
}
\hspace{0.5cm}
\subfloat[Triple expressions.]{\label{fig:grammar-texpr}%
\begin{minipage}[b]{0.45\linewidth}
  \begin{tabbing}
    \TriplePattern{} \= \hspace{0.5em} \=  \kill
    \TExpr{} \> $\coloncolonequals$ \TriplePattern{} $\mid$ {\sf `\EMPTY'} \\
    \>\> $\mid$ \SomeOfExpr{} $\mid$ \EachOfExpr{} \\
    \>\> $\mid$ \RepetExpr\\
    \TriplePattern \> $\coloncolonequals$ \PropSet{}\; {\sf '@'}\ShapeRef\\
    \SomeOfExpr \> $\coloncolonequals$ \TExpr{} {\sf `$\mid$'} \TExpr\\
    \EachOfExpr \> $\coloncolonequals$ \TExpr{} {\sf `;'} \TExpr\\
    \RepetExpr \> $\coloncolonequals$ \TExpr{} {\sf `['} $min$ {\sf `;'} $max$ {\sf `]'}\\
    \PropSet \> $\coloncolonequals$ a subset of \IRI\\
    \ShapeRef \> $\coloncolonequals$ a shape label in \SLabels
    \end{tabbing}
\end{minipage}
}
\caption{The grammar for shape expressions and triple expressions.}
\label{fig:syntax-shape-expr}
\end{figure}

\paragraph{Triple expressions.}
\label{sec:triple-expressions}
Triple expressions describe the expected neighbourhood of a node.
They are inspired by regular expressions likewise \textsc{dtd}s and \xmlschema for \xml.
A triple expression will be matched by the neighbourhood of a node in a graph, similarly to type definitions in \xmlschema that are matched by the children of some node.
The main difference is that the neighbourhood of a node in an \rdf graph is a (unordered) set, whereas the children of a node in an \xml document form a sequence.

The grammar for triple expressions (\TExpr) is given on Fig.~\ref{fig:grammar-texpr}, in which $\textit{min}$ is a natural, and $\textit{max}$ is a natural or the special value $*$.
The basic triple expression is a triple pattern and it constrains triples.
A triple expression composed of each-of (separated by a `;'), some-of (separated by a `$\mid$') and repetition operators is satisfied if some distribution of the triples in the neighborhood of a node exactly satisfies the expression. Sect.~\ref{sec:typing} defines this and draws the analogy with regular expressions.
In examples, we omit the braces for singleton \PropSet{}s, \eg we write {\sf foaf:name @\ex{StringValue}} instead of {\sf $\{$foaf:name$\}$ @\ex{StringValue}}.

\begin{example}[Shape expressions, triple expressions]
  % In schema $\sch_0$ from Sect.~\ref{sec:prelim} (and in the forthcoming examples) we write $L \to S$ as short for $\ddef(L) = S$.
  In schema $\sch_0$ from the introductory example, the definitions of the five shapes with name {\sf \ldots{}Shape\ldots} are triple expressions and collectively make use of all the operators: each-of (;), some-of ($\mid$), repetition.
  % For better readability, in $\sch_0$ we represent singleton property sets of triple patterns by the corresponding value, 
  % \eg{} {\sf foaf:name @\ex{StringValue}} stands for the \TriplePattern whose \PropSet{} is the singleton $\{\text{\sf foaf:name}\}$.
  All shapes with name {\sf \ldots{}Value\ldots} are defined by atomic \ValueDescr{}s.
  The definition of \ex{ClientAndUser} is a \ShapeAnd{} expression.\qed
\end{example}

\paragraph{Relationship between Shapes Schemas and ShEx~2.0.}
Shapes schemas slightly generalizes \shextwo and thus allows for a more concise definition of syntax and semantics.
For readers familiar with \shextwo we now explain how shapes schemas differ from \shextwo.
First, \TriplePattern{} uses a set of properties, whereas the analogous triple constraint in \shextwo uses a single property.
This slight generalization allows to encode the {\sc closed} and {\sc extra} constructs of \shextwo.
In shapes schemas, triple expressions are always closed (whereas in \shextwo they are non closed by default) but an expression $E$ can be made non-closed by transforming it into $E ; P\ @\ex{AnythgShape}[0;*]$, where $P$ is the set of all \IRI{s} not mentioned as properties in $E$, and \ex{AnythgShape} is as defined in the introductory example.
The \textsc{extra} modifier is encoded in a similar way, using sets of properties in triple patterns and negation.
Second, a \ValueDescr{} is an arbitrary set of values that can be \iri{}s, literals or blank nodes, whereas the analogous node constraint in \shextwo defines a set of allowed values using a combination of elementary constraints such as {\sc xsd} datatypes, facets, numerical intervals, node kinds.
Using an arbitrary set of values allows to get rid of unnecessary (w.r.t. defining the semantics) details.
Third, \shextwo allows to use shape labels in shape definitions; this is syntactic sugar and is equivalent to replacing the label by its definition.
Finally, in shapes schemas we omit inverse properties which would make the proofs longer without representing any additional challenge w.r.t. the semantics.

%%% Local Variables: 
%%% mode: latex
%%% TeX-master: "paper"
%%% End: 

\section{Semantics of Shapes Schemas}
\label{sec:semantics}
A shape defines the structure of a graph when visited starting from a node that has that shape.
In this section we give a precise meaning of the following statement
\vspace{-0.1cm}
\begin{center}
\MQ: node $n$ in graph $\graph$ has shape (or type) $L$ from schema $\sch$\footnote{``type'' is used as synonym of ``shape'', esp. in the notion of \emph{typing} to be introduced shortly. The use of ``type'' must not be confused with \codeil{rdf:type} from \rdfschema{}. Shapes schemas are totally independent from the \rdfschema{} vocabulary.}
\end{center}
\vspace{-0.1cm}
To give a sound definition for \MQ{} is not trivial because of the presence of recursion.
It also requires to make a design choice that we explain now.
\begin{example}[Simple recursive schema]
  \label{ex:positive-recursion}
  Let schema $\sch_1$ and graph $\graph_1$ be:
  \vspace{-0.5cm}
  \begin{code}
    \begin{tabbing}
      \ex{IssueSh} \= $\to$ \= \hspace{7,2cm} \kill \\
      \ex{IssueSh} \> $\to$ is:reportedBy @\ex{Str} ; is:relatedTo @\ex{IssueSh} [0;*] \\
      \ex{Str} \> $\to$ xsd:string\\
      \\
      \ex{i1} is:reportedBy ``Ren'' ; is:relatedTo \ex{i2} .\\
      \ex{i2} is:reportedBy ``Bob'' ; is:relatedTo \ex{i1} .
    \end{tabbing}
  \end{code}
\end{example}
Ex.~\ref{ex:positive-recursion} captures the essence of recursion.
If \codeil{\ex{i1}} has shape \codeil{\ex{IssueSh}} then \codeil{\ex{i2}} also has shape \codeil{\ex{IssueSh}}.
If on the other hand \codeil{\ex{i1}} does not have shape \codeil{\ex{IssueSh}}, then neither does \codeil{\ex{i2}}. 
This illustrates two important aspects of the semantics of shapes schemas.
First, whether a node has some shape cannot be defined independently of the shapes of the other nodes in the graph.
The consequence of this apparently simple fact is that we need a global statement about which nodes satisfy which shapes; we call this a typing.
A typing must be correct, i.e. coherent with itself.
Second, in the above example there is a (design) choice to make.
Clearly, there are two acceptable alternatives: either (1) both \codeil{\ex{i1}} and \codeil{\ex{i2}} have shape \codeil{\ex{IssueSh}}, or (2) none of them does.
Such choice is well known for recursive languages: (1) corresponds to a maximal solution, and (2) to a minimal solution.
Both choices can lead to sound semantics.
In shapes schemas we choose the maximal solution.
This is justified by applications: in the above example we \emph{do want} to consider \codeil{\ex{i1}} as a valid \codeil{\ex{IssueSh}}.
It would not be the case with semantics based on a minimal solution.

\subsection{Typing and Correct Typing}
\label{sec:typing}
The semantics is based on the notion of \emph{typing}: this is a set of couples that associate a node of an RDF graph with a shape label (a type).
In the sequel we consider a graph $\graph$ and a schema $\sch = (\SLabels, \ddef)$.
\begin{definition}[node-type association, typing]
  A \emph{node-type association} is a couple $(n,L)$ in $\nodes(\graph) \times \SLabels$.
  A \emph{typing of $\graph$ by $\sch$} is a set of node-type associations.
\end{definition}

\begin{example}
  \label{ex:typing}
  With $\sch_1$ and $\graph_1$ from Ex.~\ref{ex:positive-recursion}, the following are typings
\begin{align*}
\typing_1 &= \{\codeil{(\ex{i1}, \ex{IssueSh})}, \codeil{(\ex{i2}, \ex{IssueSh})}, \codeil{("Ren", \ex{Str})}, \codeil{(``Bob'', \ex{Str})}\}\\
\typing_2 &= \{\codeil{(``Ren'', \ex{Str})}, \codeil{(``Bob'', \ex{Str})}\}\\
\typing_3 &= \{\codeil{(\ex{i1}, \ex{IssueSh})}, \codeil{(\ex{i2}, \ex{IssueSh})}\}\\
\typing_4 &= \emptyset.
\end{align*}
\end{example}
A typing is correct if, intuitively, it contains an evidence for every node-type association in it. 
In the above example $\typing_1$ and $\typing_2$ are correct, whereas $\typing_3$ is not correct as it contains e.g. the association $\codeil{(\ex{i1}, \ex{IssueSh})}$ but does not contain the association $\codeil{("Ren", \ex{Str})}$ that is required for $\codeil{\ex{i1}}$ to have type $\codeil{\ex{IssueSh}}$.
The empty typing ($\typing_4$) is always correct.

\begin{definition}[correct typing]
  \label{def:correct-typing}
  Let $\typing \subseteq \nodes(\graph) \times \sch$.
  We say that \emph{$\typing$ is a correct typing} if for any $(n, L) \in \typing$, it holds $\typing, n \sat \ddef(L)$, where $\sat$ is the relation defined on Fig.~\ref{fig:semantics-shapes}.
\end{definition}

\begin{figure}[t]
\subfloat[Node satisfies a shape expression.]{\label{fig:semantics-shapes}
  \begin{minipage}[b]{0.45\linewidth}
    \input{shexpr-sat}
  \end{minipage}
}
\hspace{0.5cm}
\subfloat[Set of triples matches a triple expression.]{\label{fig:semantics-texpr}
  \begin{minipage}[b]{0.45\linewidth}
    \input{texpr-sat}
  \end{minipage}

}
  \caption{Definitions of the $\sat$ and $\match$ relations.}
  \label{fig:semantics-shapes-texpr}
\end{figure}

\paragraph{Discussion on $\sat$.} 
For a shape expression $S$, the definition of $\typing, n \sat S$ on Fig.~\ref{fig:semantics-shapes} is by recursion on the structure of $S$.
In Rules~\sevaluedescr, $V$ is a subset of $\IRI \cup \Lit \cup \{\blank\}$ defining a \ValueDescr.
A node $n$ satisfies the value description $V$ if $n$ belongs to the set $V$, or if $n$ is a blank node and \blank{} is in $V$.
The other base case is Rule~\seneighbdescr, in which $E$ is a $\TExpr$ representing a neighbourhood description. 
A node $n$ satisfies the \NeighbDescr{} $E$ if the neighbourhood of $n$ matches the triple expression $E$.
The matching relation $\match$ is defined on Fig.~\ref{fig:semantics-texpr} and discussed below.
The remaining four rules are for the Boolean operators.
The rules for $\AND$ and $\OR$ are as one would expect.
Regarding negation, a node satisfies a $\ShapeNot$ expression if it does not satisfy its sub-expression, as stated by Rule~\seshapenot.
The premise of that rule is $\typing, n \not\sat S$ and means that (using the inference rules on Fig.~\ref{fig:semantics-shapes}) it is impossible to construct a proof for $\typing,n \sat S$.

\paragraph{Discussion on $\match$.}
For a set of triples $N$, a typing $\typing$ and a $\TExpr$ $E$, we say that \emph{$N$ matches $E$ with $\typing$}, and we write $\typing, N \match E$, as defined recursively on the structure of $E$ on Fig.~\ref{fig:semantics-texpr}.
Note that the $\match$ relation is defined for an arbitrary set of triples $N$. 
In practice, $N$ will be (a subset of) the neighbourhood of some node.
In the basic Rule~\tetriplepattern, $P@L$ is a \TriplePattern with $P$ a set of \IRI{}s and $L$ a shape label.
A singleton set of triples $\{(\mathit{subj}, \mathit{pred}, \mathit{obj})\}$ matches the triple pattern if the predicate $\mathit{pred}$ belongs to $P$ and the object has type $L$ in $\typing$.
The other basic rule is Rule~\teempty: an empty set of triples satisfies the $\EMPTY$ triple expression.

The remaining rules are about the composed triple expressions.
A set of triples matches a \SomeOfExpr if it matches one of its sub-expressions (Rules~\tesomeof).
The semantics of a \EachOfExpr is a bit more complex.
A set $N$ matches an each-of triple expression $E_1 ; E_2$ if $N$ is the disjoint union of two sets $N_1$ and $N_2$, and $N_1$ matches the sub-expressions $E_1$, and $N_2$ matches the sub-expression $E_2$.
Let us make a parallel between regular expressions and triple expressions.
The each-of operator is analogous to concatenation.
Recall that a string $w$ matches a regular expression $R_1\cdot R_2$ (where $\cdot$ is concatenation) whenever $w$ can be ``split'' into two strings $w_1$ and $w_2$ such that their concatenation gives $w$ ($w = w_1\cdot w_2$), and $w_1$ matches $R_1$, and $w_2$ matches $R_2$.
In the case of triple expressions, the set of triples $N$ is "split" into two disjoint sets $N_1$ and $N_2$: disjoint union on sets is analogous to concatenation on words.
Following the same analogy, repetition in triple expressions corresponds to Kleene star (the star operator) in regular expressions, with the difference that it allows to express arbitrary intervals for the number of allowed repetitions, whereas Kleene star is always $[0,*]$.
So, in Rule~\terepet, a set of triples $N$ matches a repetition triple expression $E[\mathit{min};\mathit{max}]$ if $N$ can be split as the disjoint union of $k$ sets $N_1, \ldots, N_k$ such that $k$ is within the interval bound $[\mathit{min};\mathit{max}]$ and each of these sets matches the sub-expression $E$.
Note that $k = 0$ is possible only when $N = \emptyset$.

\bigskip
The laws of the Boole algebra can be used to put a shape expression in disjunctive normal form in which only atomic sub-expressions \ValueDescr{} and \NeighbDescr{} are negated.
From now on we consider only shape expressions in disjunctive normal form.
Note also that the each-of and some-of operators are associative and commutative and we use them as operators of arbitrary arity, as \eg{} in schema $\sch_0$ from the introductory example.

\subsection{Stratified Negation}
\label{sec:restricted-negation}

Because of the presence of recursion and negation, the notion of correct typing is not sufficient for defining sound semantics of shapes schemas.

\begin{example}[Negation and recursion]
\label{ex:negative-recursion}
Let schema $\sch_2$ and graph $\graph_2$ below:
\begin{center}
  \begin{code}
    \begin{minipage}{0.45\linewidth}
      \ex{L1} $\to$ \NOT(\codeil{ex:p} \ex{L2})\\
      \ex{L2} $\to$ \NOT(\codeil{ex:p} \ex{L1})
    \end{minipage}
    \begin{minipage}{0.45\linewidth}
      \ex{n1} ex:p \ex{n2} .\\
      \ex{n2} ex:p \ex{n1} . 
    \end{minipage}
  \end{code}
\end{center}
These two typings of $\graph_2$ by $\sch_2$ are both correct: $\typing_5 = \{(\codeil{\ex{n1}, \ex{L1}})\}$ and $\typing_6 = \{(\codeil{\ex{n2}, \ex{L2}})\}\hspace{2cm}$\qed
\end{example}
The two typings in Ex.~\ref{ex:negative-recursion} strongly contradict each other.
In order to prove that node \codeil{\ex{n1}} has shape \codeil{\ex{L1}} (in $\typing_5$), we need to prove that \codeil{\ex{n1}} does \textbf{not} have shape \codeil{\ex{L2}}.
The latter however does hold in $\typing_6$.
Such strong contradictions are possible only in presence of negation.
In comparison, in Ex.~\ref{ex:positive-recursion} we also have two contradicting typings, but none of them uses in its proof a negative statement that is positive in the other typing.

This problem is well known in logic programming \eg{} in Datalog, see Chapter~15 in~\cite{alice-book} for an overview.
The literature considers several solutions for defining coherent semantics in this case, among which the most popular are negation-as-failure, stratified negation and well-founded semantics.
For instance, well-founded semantics would answer \emph{undefined} to the question ``does $n$ have shape $L$'' whenever there exist two proofs that contradict each-other on that fact.
We exclude this solution for two reasons: it is not helpful for users, and it might require to compute all possible typings which is costly.
We opt for stratification semantics instead.
It imposes a syntactic restriction on the use of recursion together with negation, so that schemas as the one on Ex.~\ref{ex:negative-recursion} are not allowed.
This is a reasonable restriction because negation in ShEx is expected to be used mainly locally, \eg{} to forbid some property in the neighbourhood of a node.

We now define of stratified negation. 
The \emph{dependency graph} of $\sch$ is a graph whose set of nodes is $\SLabels$, and that has two kinds of edges labelled $\depneg$ and $\deppos$ defined by (recall that shape expressions in disjunctive normal form):
\begin{itemize}
\item There is a \emph{negative dependency} edge $\depneg(L_1, L_2)$ from $L_1$ to $L_2$ iff the shape label $L_2$ appears in $\ddef(L_1)$ under an occurrence of the $\NOT$ operator;
\item There is a \emph{positive dependency} edge $\deppos(L_1, L_2)$ from $L_1$ to $L_2$ iff the shape label $L_2$ appears in $\ddef(L_1)$ but never under an occurrence of $\NOT$.
\end{itemize}
\begin{definition}[schema with stratified negation]
A schema $\sch = (\SLabels, \ddef)$ is \emph{with stratified negation} if there exists a natural number $k$ and a mapping $\strat$ from $\SLabels$ to the interval $[1;k]$ such that for all shape labels $L_1, L_2$:
\begin{itemize}
\item if $\depneg(L_1, L_2)$, then $\strat(L_1) > \mathit{strat}(L_2)$;
\item if $\deppos(L_1, L_2)$, then $\strat(L_1) \ge \mathit{strat}(L_2)$.
\end{itemize}
\end{definition}
The mapping $\strat$ is called a \emph{stratification} of $\sch$.
A well known property of stratified negation is that the dependency graph does not have a cycle that goes through a negative dependency edge.
This intuitively means that if shape $L_1$ depends negatively on shape $L_2$, then $L_2$ does not (transitively) depend on $L_1$.
Positive interdependence is allowed in an unrestricted manner, as in $\sch_1$ from Ex.~\ref{ex:positive-recursion}.
$\sch_2$ from Ex.~\ref{ex:negative-recursion} is not with stratified negation because $\depneg(\codeil{\ex{L1}, \ex{L2}})$ and $\depneg(\codeil{\ex{L2}, \ex{L1}})$

\begin{example}[Stratification]
  \label{ex:stratification}
  Let schema $\sch_3$ below.
  \begin{center}
    \begin{code}
      \begin{minipage}[t]{0.63\linewidth}
        \ex{L1} $\to$ \NOT(\codeil{ex:a} @\ex{L2} ; \codeil{ex:b} @\ex{Str})\\
        \ex{L2} $\to$ \codeil{ex:c} @\ex{L3}\\
      \end{minipage}
      \begin{minipage}[t]{0.33\linewidth}
        \ex{L3} $\to$ \codeil{ex:c} @\ex{L2}\\
        \ex{Str} $\to$ \codeil{xsd:string}        
      \end{minipage}
    \end{code}
  \end{center}
  The dependency graph contains the edges $\depneg(\ex{L1},\ex{L2})$, $\depneg(\ex{L1},\ex{Str})$, $\deppos(\ex{L2},\ex{L3})$, $\deppos(\ex{L3},\ex{L2})$.
  The unique loop is around \ex{L2} and \ex{L3} and it goes through positive dependencies only, so the schema is stratified.
  A stratification should be such that \ex{Str} and \ex{L2} are on stratums strictly lower than \ex{L1}, and \ex{L2} and \ex{L3} are on the same stratum.
  One possible stratification is \ex{L1} on stratum 2 and the other three shape labels on stratum 1.
  Another one is \ex{L2} and \ex{L3} on stratum 1, \ex{Str} on stratum 2, and \ex{L1} on stratum 3. 
  The latter is called a most refined stratification as none of the stratums can be split.
\end{example}

\subsection{Maximal Correct Typing}
Recall from Ex.~\ref{ex:typing} that both $\typing_1$ and $\typing_4$ are correct.
Note that $\codeil{\ex{i1}}$ has shape $\codeil{\ex{IssueSh}}$ according to $\typing_1$ but not according to $\typing_4$.
Then what is the correct answer of \MQ{} for $\codeil{\ex{i1}}$ and $\codeil{\ex{IssueSh}}$?
Does $\codeil{\ex{i1}}$ have shape \ex{IssueSh} at the end?
This section provides an answer to that question.
In one sentence: we trust $\typing_1$ because it is greater; actually it is the greatest (maximal) typing.
The comparison is based on set inclusion.

The following Lemma~\ref{lem:greatest-typing-neg-free} establishes that a maximal typing always exists in absence of negation. 
The proof is based on Lemma~2 in \cite{icdt2015} that can be easily extended for the richer schemas we have here.
\begin{lemma}
  \label{lem:greatest-typing-neg-free}
  Let $\sch$ be a schema that does not use the negation operator $\NOT$.
  Then for all graphs $\graph$, there exists a correct typing $\typing_{\mathrm{g}}$ of $\graph$ by $\sch$ such that for every $\typing'$, if $\typing'$ is a correct typing of $\graph$ by $\sch$, then $\typing' \subseteq \typing_{\mathrm{g}}$.
\end{lemma}
The typing $\typing_{\mathrm{g}}$ can be computed as the union of all correct typings $\typing'$.

Let us now define a maximal typing in presence of negation.
Let $\strat$ be a stratification of $\sch$ that has $k$ strata, with $k \ge 1$.
For any $1 \le i \le k$, the schema $\sch_i$ is the restriction of $\sch$ that uses only the shape labels whose stratum is less than $i$.
Formally, $\sch_i = (\SLabels_i, \ddef_i)$ with $\SLabels_i =  \{L \in \SLabels \mid \strat(L) \le i\}$, and their respective definitions $\ddef_i(L) = \ddef(L)$.
Remark that if $\sch$ is stratified, then $\sch_1$ is negation-free.

For a set of labels $\SLabels_i \subseteq \SLabels$, $\typing_{|\SLabels_i} = \{(n,L) \in \typing \mid L \in \SLabels_i\}$ is the restriction of $\typing$ on the labels from $\SLabels_i$.
\begin{definition}[stratification-maximal correct typing]
  \label{def:stratified-maximal-correct-typing}
  Let $\sch = (\SLabels, \ddef)$ be a schema, $\graph$ be a graph, and $\strat$ be a stratification of $\sch$ with $k$ stratums (for $k \ge 1$).
  For any $1 \le i \le k$, let $\typing_i$ be the typing of $\graph$ by $\sch_i$, defined by:
  \begin{itemize} 
  \item $\typing_1$ is the maximal correct typing of $\graph$ by $\sch_1$, as defined in Lemma~\ref{lem:greatest-typing-neg-free};
  \item for any $1 \le i < k$, $\typing_{i+1}$ is the union of all correct typings $\typing'$ of $\graph$ by $\sch_{i+1}$ s.t. $\typing'_{|\SLabels_i} = \typing_i$.
  \end{itemize}
  The \emph{stratification-maximal correct typing of $\graph$ by $\sch$ with stratification $\strat$} is $Typing(\graph, \sch, \strat) = \typing_k$.
\end{definition}
$\Typing(\graph, \sch, \strat)$ from the above definition is indeed a correct typing for $\graph$ by $\sch$, as shown in the following proposition that is the core of the proof of soundness for the semantics of shapes schemas.
\begin{proposition}
  \label{prop:Typing-is-correct}
  For any schema $\sch$, any stratification $\strat$ of $\sch$ and any graph $\graph$, $\Typing(\graph, \sch, \strat)$ is a correct typing of $\graph$ by $\sch$.
\end{proposition}
\begin{proof}
  Goes by induction on the number of stratums. 
  The base case (1 stratum) is Lemma~\ref{lem:greatest-typing-neg-free}.
  For the induction case and stratum $i+1$, by induction hypothesis $\typing_i$ is correct for $\graph$ and $\sch_i$.
  It is enough to show that if $\typing'$ and $\typing''$ are two correct typings for $\graph$ by $\sch_{i+1}$ and $\typing'_{|\SLabels{i}} = \typing''_{|\SLabels{i}} = \typing_i$, then their union $\typing = \typing' \cup \typing''$ is correct for $\graph$ by $\sch_{i+1}$.
  Let $(n,L) \in \typing$ and suppose that $(n,L) \in \typing'$. 
  Because $\typing'$ is correct, we have $\typing', n \sat \ddef(L)$. 
  We will show that (*) the proof for $\typing', n \sat \ddef(L)$ can be used as a proof for $\typing, n \sat \ddef(L)$.
  If $\ddef(L)$ does not contain a negation of a triple expression, then (*) easily follows from the definition of $\sat$.

  So suppose $\ddef(L)$ contains a negation operator on top of the triple expressions $E_1, \ldots, E_l$.
  That is, (recall that shape expressions are in disjunctive normal form), $\NOT E_j$ is a sub-expression of $\ddef(L)$ for every $1 \le j \le l$.
  Then the proof for $\typing', n \sat \ddef(L)$ contains applications of Rule~\seshapenot for $\NOT E_j$ that witness that there does not exist a proof for $\typing', n \sat E_j$, for every $1 \le j \le l$.
  We need to show that a proof $\typing, n \sat E_j$ cannot exist.
  Suppose by contradiction that $P$ is a proof for $\typing, n \sat E_j$, for some $1 \le j \le l$.
  Let $\SLabels'$ be the set of all shape labels that appear in $E_j$, then $P$ uses only node-type associations with labels from $\SLabels'$.
  That is, $\typing_{|\SLabels'}, n \sat E_j$ holds.
  As $E_j$ is negated in $\ddef(L)$, we have $\SLabels' \subseteq \SLabels_i$, so $\typing_{|\SLabels_i}, n \sat E_j$ also holds.
  But $\typing_{|\SLabels_i} = \typing_i \subseteq \typing'$.
  Contradiction.\qed
\end{proof}
 
Lemma~\ref{lem:maximal-typing-unique-stratification} below establishes that $Typing(\graph, \sch, \strat)$ does not depend on the stratification being chosen.
This allows to define the maximal correct typing (Def.~\ref{def:maximal-correct-typing}) and to give a precise meaning of \MQ{} (Def.~\ref{def:mq}) which was the objective of this section.

\begin{lemma}
  \label{lem:maximal-typing-unique-stratification}
  Let $\sch = (\SLabels, \ddef)$ be a schema and $\graph$ be a graph.
  Let $\strat_1$ and $\strat_2$ be two stratifications of $\sch$.
  Then $Typing(\graph, \sch, \strat_1) = Typing(\graph, \sch, \strat_2)$.
\end{lemma}
\begin{proof}(Idea)
  The proof uses a classical technique as \eg for stratified Datalog.
  There exists a unique (up to permutation on the numbering of stratums) most refined stratification $\strat_{\text{ref}}$ such that for any other stratification $\strat'$, each stratum of $\strat'$ can be obtained as a union of stratums of $\strat_{\text{ref}}$.
  Then we show that for any stratification $\strat'$, $Typing(\graph, \sch, \strat') = Typing(\graph, \sch, \strat_{\text{ref}})$.
\end{proof}
\begin{definition}[maximal correct typing]
  \label{def:maximal-correct-typing}
  Let $\sch = (\SLabels, \ddef)$ be a schema and $\graph$ be a graph.
  The \emph{maximal correct typing of $\graph$ by $\sch$} is denoted $Typing(\graph, \sch)$ and is defined as $Typing(\graph, \sch, \strat)$ for some stratification $\strat$ of $\sch$.
\end{definition}
\begin{definition}[\MQ]
  \label{def:mq}
  Let $\sch = (\SLabels, \ddef)$ be a schema and $\graph$ be a graph.
  We say that node $n$ (of $\graph$) has shape $L$ (from $\sch$) if $(n,L) \in Typing(\graph,\sch)$.
\end{definition}

\section{Validation}
\label{sec:compute}
In Sect.~\ref{sec:semantics} we have given a declarative semantics of the shapes language.
We now consider the related computational problem.
We are again interested by the \MQ{} statement (as defined in Sect.~\ref{sec:semantics}), \ie{} checking whether a given node has a given shape.

\subsection{Refinement Algorithm}
\label{sec:refinement-algorithm}
Algorithm~\ref{algo:refinement-algorithm} computes $\Typing(\graph, \sch, \strat)$.
\begin{algorithm}[t]
  \KwIn{$\graph$: a graph, $\sch = (\SLabels, \ddef)$: a schema, $\strat$ a stratification for $\sch$ with $k$ strata}
  \KwOut{$\Typing(\graph, \sch)$}
  \BlankLine

  $\typing \gets \emptyset$\;
  \For{$i$ from $1$ to $k$}{\nllabel{line:refine-main-loop}
    \tcp{Add all node-type associations for stratum $i$}
    \ForEach{$n$ in $\nodes(\graph)$}{
      \ForEach{$L$ in $\SLabels_i$}{
        add $(n,L)$ to $\typing$\;
      }
    }
    \tcp{Refine w.r.t the types on stratum $i$}
    $\mathit{changing} \gets \true$\;
    \While{$\mathit{changing}$}{
      $\mathit{changing} \gets \false$\;
      \ForEach{$(n,L)$ in $\typing$ s.t. $L \in \SLabels_i$}{
        \If{not $\typing, n \sat \ddef(L)$}{
          remove $(n,L)$ from $\typing$\;\nllabel{line:refinement-remove}
          $\mathit{changing} \gets \mathsf{true}$\;
        }
      }
    }
  }
  \Return{$\typing$}
  \caption{The algorithm $\refinealgo(\graph, \sch, \strat)$.\label{algo:refinement-algorithm}}
\end{algorithm}
The $i$-th iteration of the loop on line~\ref{line:refine-main-loop} computes $\typing_i$ from Def.~\ref{def:stratified-maximal-correct-typing}.
The algorithm is correct thanks to Lemma~2 from \cite{icdt2015} applied to every stratum $i$.
According to that lemma, the maximal typing defined as the union of all correct typings (\ie{} $\typing_i$) can be computed by iteratively removing unsatisfied node-type associations (done on line~\ref{line:refinement-remove}) until a fixed point is reached (detected when $\mathit{changing}$ remains $\false$).
The advantage of the $\refinealgo$ algorithm is that once $\Typing(\graph, \sch)$ is computed, testing whether node $n$ has shape $L$ is done with no additional cost by testing whether $(n,L)$ belongs to $\Typing(\graph,\sch)$.
The drawback is that it considers all node-type associations which is not always necessary, as shown here after.

\vspace{-0.3cm}
\subsection{Recursive Algorithm}
\label{sec:recursive-algorithm}

Algorithm~\ref{algo:recalgo} allows to check whether node $n$ has shape $L$ without constructing $\Typing(\graph,\sch)$. 
The idea is to visit only a sufficiently large portion of $\Typing(\graph,\sch)$.
\begin{example}[Motivation of the \recalgo{} algorithm]
  Considering schema $\sch_3$ from Ex.~\ref{ex:stratification} and graph $\graph_3$ below:
  \label{ex:illustration-recalgo}
  \begin{center}
    \begin{code}
      \begin{minipage}[t]{0.4\linewidth}
        \codeil{ex:n1} \codeil{ex:a} \codeil{ex:n2} \\
        \codeil{ex:n1} \codeil{ex:b} 4 .
      \end{minipage}
      \begin{minipage}[t]{0.4\linewidth}
        \codeil{ex:n2} \codeil{ex:c} \codeil{ex:n3} .\\
        \codeil{ex:n3} \codeil{ex:c} \codeil{ex:n2} .
      \end{minipage}
    \end{code}
  \end{center}
  We want to check whether \codeil{ex:n1} has shape \ex{L1}.
  Remark that the neighbor nodes of \codeil{ex:n1} are \codeil{ex:n2} and 4, whereas the shape labels on which the definition of \ex{L1} depends are \ex{L2} and \ex{Str}.
  Any correct proof for $\typing, \codeil{ex:n1} \sat \ex{L1}$ (or for $\typing, \codeil{ex:n1} \not\sat \ex{L1}$) would have as leaves either applications of Rule~\sevaluedescr that do not depend on $\typing$, or applications of Rule~\tetriplepattern that uses node-type associations $(n,L)$ where $n$ is a neighbor of \codeil{ex:n1} and $L'$ is a label such that $\deppos(\ex{L1},L')$ or $\depneg(\ex{L1},L')$.
\end{example}

Assume schema $\sch = (\SLabels, \ddef)$ and graph $\graph$.
For a shape label $L$ in $\sch$ and a node $n$ in $\graph$, we denote $\deps(n, L)$ the set of node-type associations $(n', L')$ s.t. $n'$ is a neighbor of $n$ (that is, $(n, p, n') \in \neigh(n)$ for some \IRI{} $p$) and $L'$ appears as a shape reference in $\ddef(L)$.
Algorithm~\ref{algo:recalgo} uses this easy to show property: $\typing, n \models \ddef(L)$ iff $\typing \cap \deps(n,L), n \models \ddef(L)$.
In order to check whether $n$ has shape $L$,  Algorithm~\ref{algo:recalgo} will (recursively) check whether $n'$ has shape $L'$ for all $(n',L')$ in $\deps(n,L)$.
The parameter $\Hyp$ is a stack of node-type associations that is also seen (on line~\ref{line:recalgo-loop}) as the set of node-type associations it contains.
$\Dep$ is a set of node-type associations.
\begin{algorithm}[t]
  \SetKw{KwBreak}{break}
  \KwIn{$n$: node in $\graph$, $L$: label in $\SLabels$, $\Hyp$: a stack over $\nodes(G) \times \SLabels$}
  \KwOut{$\true$ if $n$ has label $L$, $\false$ otherwise}
  \BlankLine
  \nllabel{line:recalgo-push}
  $\Hyp = \Hyp.\push((n,L))$\;
  $\Dep = \emptyset$\;\nllabel{line:dep-decl}
  \ForEach{$(n',L')$ in $\deps(n,L) \smallsetminus \Hyp$ \nllabel{line:recalgo-loop}}{
      \If{$\recalgo(n',L', \Hyp)$}{\nllabel{line:recalgo-reccall}
        $\Dep = \Dep \cup \{(n', L')\}$\;
      }
  }
  result = $\Dep \cup \Hyp, n \sat \ddef(L)$ \nllabel{line:recalgo-result}\;
  $\Hyp = \Hyp.\pop()$\;
  \Return result \;\nllabel{line:recalgo-last-return} 
  \caption{$\recalgo(n, L, \Hyp)$\label{algo:recalgo}. Graph $\graph$ and schema $\sch$ are global variables.}
\end{algorithm}

\begin{example}[Execution trace of the $\recalgo$ algorithm]
  \label{ex:recalgo-execution-trace}
  Here is the tree of recursive calls generated during the evaluation of $\recalgo(\codeil{ex:n1}, \ex{L1}, [])$ for graph $\graph_3$ and schema $\sch_3$, where $[]$ is the empty stack. The returned value is given on the right.
  $\recalgo(\codeil{ex:n1}, \ex{L1}, [])$ generates four recursive calls that correspond to $\deps(\codeil{ex:n1}, \ex{L1})$.
  The call for $\codeil{ex:n3}$ and $\ex{L3}$ does not generate any recursive call: $\deps(\codeil{ex:n3}, \ex{L3})$ contains only $(\codeil{ex:n2},\ex{L2})$ which is on the stack.

\newcommand{\stepzeroend}{$\cdot$}
\newcommand{\stepone}{$|$--}
\newcommand{\steptwo}{$|$~$|$--}
\smallskip\noindent
\begin{tabular}{l|l}
  $\recalgo(\codeil{ex:n1}, \ex{L1}, [])$ & ~$\true$\\
  \stepone$\recalgo(\codeil{ex:n2}, \ex{L2}, [(\codeil{ex:n1},\ex{L1})])$ & ~$\true$\\
  \steptwo$\recalgo(\codeil{ex:n3}, \ex{L3}, [(\codeil{ex:n1},\ex{L1}), (\codeil{ex:n2},\ex{L2})])$~~~ & ~$\true$\\
  \stepone$\recalgo(\codeil{ex:n2}, \ex{Str}, [(\codeil{ex:n1},\ex{L1})])$ & ~$\false$\\
  \stepone$\recalgo(4, \ex{L2}, [(\codeil{ex:n1},\ex{L1})])$ & ~$\false$\\
  \stepone$\recalgo(4, \ex{Str}, [(\codeil{ex:n1},\ex{L1})])$ & ~$\false$
\end{tabular}
\end{example}

The correctness of the $\recalgo$ algorithm is stated by the following:
\begin{proposition}[Correctness of the $\recalgo$ algorithm]
 \label{thm:recalgo-correct}
 For any node $n$ and any shape label $L$, the evaluation of $\recalgo(n, L, [])$ terminates and returns true if $(n,L) \in \Typing(\graph, \sch)$ and false otherwise.
\end{proposition}
\begin{proof}[Sketch]
  For termination: the recursion cannot be infinite-breadth as $\recalgo$ generates a finite number of recursive calls on line~\ref{line:recalgo-reccall}.
  Infinite-depth recursion is also impossible because $\Hyp$ is a call stack and the condition on line~\ref{line:recalgo-loop} prevents from (recursively) calling $\recalgo$ with the same node and label.

  The proof of correctness goes by induction on the stratum of $L$ using the most refined stratification $\strat$.
  For every stratum $i$ we show that whenever $\Hyp$ contains only node-type associations $(n',L')$ with  $strat(L') > i$, and for any $L$ s.t. $\strat(L) = i$, $\Typing, n \sat \ddef(L)$ iff $\recalgo(n,L,\Hyp)$ returns true.
  For the $\Rightarrow$ direction, the main argument is that if $\Typing, n \sat \ddef(L)$ then also $\Typing \cup \Hyp, n \sat \ddef(L)$.
  This is not true in general because of negation, but is true if $\Hyp$ is on stratum $\ge \strat(L)$ as in this case no type in $\Hyp$ is negated in $\ddef(L)$.
  For the $\Leftarrow$ direction, we need to show that if $\recalgo(n,L,\Hyp)$ returns $\true$ then $(n,L) \in \Typing(\graph,\sch)$.
  The problematic case is when $\recalgo(n,L,\Hyp)$ returns $\true$ whereas $n$ does not have label $L$.
  Such error necessarily comes from the fact that on line~\ref{line:recalgo-result} the algorithm used some $(n',L') \in \Hyp \smallsetminus \Typing(\graph,\sch)$ in the proof for $\Dep \cup \Hyp, n \sat \ddef(L)$.
  Consequently, $\strat(L) = \strat(L')$, and because we consider the most refined stratification, it follows that $L$ and $L'$ mutually depend on each other in the dependency graph of $\graph$.
  Then we need to distinguish two cases.
  Either all shape labels on stratum $i$ only depend on each others, as for instance \ex{L2} and \ex{L3} from Ex.~\ref{ex:stratification}.
  In that case $\recalgo(n,L,\Hyp)$ returns $\true$ based only on hypotheses in $\Hyp$, which is correct w.r.t. the semantics based on maximal solution: if nothing outside stratum $i$ allows to disprove that $n$ has label $L$, then it is indeed the case.
  The other possibility is that a shape label $L'$ on stratum $i$ depends also on shapes from lower stratums, as \ex{IssueSh} from Ex.~\ref{ex:positive-recursion} that depends on \ex{Str}.
  Then the test on line~\ref{line:recalgo-result} of the call of $\recalgo$ with $L'$ will take this dependency into account and return $\true$ only if all conditions, including those that depend on the lower stratums, are satisfied. \qed
\end{proof}

\subsection{On Implementation of the Validation Algorithms}
\label{sec:optimization}

Both algorithms use a test for $\typing, n \sat \ddef(L)$, which non trivial part is the test of the $\match$ relation required in Rule~\seneighbdescr{}.
The latter is equivalent to checking whether a word (a string) matches a regular expression disregarding the ordering of the letters of the word.
Here the word is over the alphabet of triple patterns that occur in the triple expression.
In \cite{jose-rdf-validation} we presented an algorithm for this problem based on regular expression derivatives.
In \cite{icdt2015} we gave another algorithm for so called deterministic single-occurrence triple expressions.
That algorithm can be extended to general expressions, and was used in several of the implementations of ShEx available as open source\footnote{A list of the available ShEx implementations can be found on \url{http://shex.io/}}.

The $\recalgo$ algorithm was presented in a form that is easier to understand but not optimized.
An implementation could reduce considerably the search space of the algorithm by exploring only relevant node-shape associations from $\deps(n,L)$ 
For instance, in Ex.~\ref{ex:recalgo-execution-trace} checking 4 against \codeil{L2} is useless first because 4 is accessible from \codeil{ex:n1} by \codeil{ex:b} whereas \ex{L2} in the schema is accessible from \ex{L1} by \codeil{ex:a}.

A more involved version of the $\recalgo$ algorithm could memorize portion of $\Typing(\graph, \sch)$ to be reused.
This however should be done carefully: one should not memorize all node-shape associations $(n,L)$ for which the algorithm returned $\true$, as some of these can be false positives as discussed in the proof of Prop.~\ref{thm:recalgo-correct}.

%%% Local Variables: 
%%% mode: latex
%%% TeX-master: "paper"
%%% End: 

% \section{Related work}
% \label{sec:related}
% \input{related}

\section{Conclusion}
\label{sec:concl}
In this paper we introduced shapes schemas that formalize the semantics of \shextwo and we showed that the semantics of \shextwo is sound.
We also presented two algorithms for validating an \rdf{} graph against a shapes schema.

ShEx and the underlying formalism presented here are still evolving, and there are several promising directions some of which are already being explored: introduce operators for value comparison, use property paths in triple patterns, define an \rdf{} transformation language based on ShEx.
We also plan to consider several heuristics and optimizations as the ones discussed in Sect.~\ref{sec:optimization} in order to accelerate the validation of shapes schemas.
These will be validated on examples.
Another open problem is error reporting in ShEx: how to give useful feedback for correcting validation errors.
We also plan to explore the exact relationship between shapes schemas and \shacl and establish whether shapes schemas can be encoded in \sparql{} extended with recursion as the one defined in~\cite{Reuter2015}.

\paragraph{Acknowledgments.}
This work was partially supported by CPER Nord-Pas de Calais/FEDER DATA Advanced data science and technologies 2015-2020, ANR project DataCert ANR-15-CE39-0009.

%%% Local Variables: 
%%% mode: latex
%%% TeX-master: "paper"
%%% End: 

\bibliographystyle{plain}
\bibliography{biblio}

% \newpage
% \appendix
% \section{Proof of Correctness of the Recursive Algorithm}
% \label{app:proof-correct-recalgo}
% \input{recalgo-proof-correctness}

%\input{sec-recursive-algorithm-proof}

\end{document}